\documentclass[paper]{JHEP3}
\usepackage[centertags]{amsmath}
\usepackage{amsfonts} \usepackage{amssymb} \usepackage{amsthm}
\usepackage{graphicx}

\def\one{{\rm 1\kern -.9mm l}}                             %

\def\beq{\begin{equation}}
\def\eeq{\end{equation}}
\def\beq{\begin{equation}}
\def\eeq{\end{equation}}
\def\beqa{\begin{eqnarray}}
\def\eeqa{\end{eqnarray}}
\newcommand{\eqa}{\begin{eqnarray}}
\newcommand{\ena}{\end{eqnarray}}
\newcommand{\eq}[1]{eq. (\ref{#1})}

\def\ee{\mathrm{e}}

\newcommand{\cale}{\mathcal{E}}
\title{Universal behaviour
of interfaces in 2d and dimensional reduction of Nambu-Goto strings%
\thanks{Work partially supported by the European Community's Human Potential
Programme under contract MRTN-CT-2004-005104 ``{Constituents, Fundamental
Forces and Symmetries of the Universe}'' and by the European Commission TMR
programme HPRN-CT-2002-00325 (EUCLID)
and by the Italian M.I.U.R under contract
PRIN-2005023102 {``Strings, D-branes and Gauge Theories''}.
}}
\author{M. Bill\'o, M. Caselle, L. Ferro
\\
Dipartimento di Fisica Teorica, Universit\`a di Torino\\
and Istituto Nazionale di Fisica Nucleare - sezione di Torino \\
Via P. Giuria 1, I-10125 Torino, Italy
}
\abstract{We propose a simple effective model for the description of interfaces
in 2d statistical models, based on the first-order treatment of an action
corresponding to the length of the interface.  The universal prediction of
this model for
the interface free energy agrees with the result of an exact calculation
in the case of the 2d Ising model.
This model appears as a
dimensional reduction of the Nambu-Goto stringy description of interfaces in 3d,
i.e., of the capillary wave model. }
\preprint{DFTT/11/2007}

\begin{document}

\section{Introduction}
\label{sec:intro}
The appearance of interfaces in statistical systems under particular conditions
has always raised much interest in various fields of research, ranging
from condensed matter to high energy physics. Recently, there have been remarkable
theoretical and computational improvements in the study of this phenomenon.

Numerical investigations of interfaces in statistical systems
(and of their analogue in Lattice Gauge Theories) have attained a great level
of accuracy and reliability. In particular, the interface free energy in the
Ising 3d model has been thoroughly studied
by means of Monte Carlo
simulations. Indeed, spin models provide a simple realization of interfaces
since in their broken symmetry phase
an interface separating
coexisting vacua of different magnetization can be easily induced in the system
by suitably choosing the boundary conditions.

On the theoretical side, different effective models can be used to describe
the behaviour of interfaces in 3d systems
and in particular to evaluate their free energy.
The most popular is the {\em
capillary wave model} (CWM)\cite{BLS,rw82} which is based on the assumption
of an action proportional to the area of the surface swept by the interface
(for a review, see for instance \cite{Privman:1992zv}).
This model is tantamount to consider the interfaces as bosonic strings
embedded in three dimensions, with a Nambu-Goto
action~\cite{Goto:1971ce,Nambu:1974zg}. Although this
effective string approach neglects the conformal anomaly which appears for
target space dimensions $d\not= 26$, the effects of this
approximation appear to be subleading \cite{Olesen:1985pv} for large
worldsheets. This description has exactly the same nature of the effective
string description of certain observables, such as Polyakov and Wilson loops,
in LGT (see for instance \cite{Caselle:2005xy,Kuti:2005xg} and references
therein).

The standard procedure to treat the effective Nambu-Goto string \cite{df83}
in the last twenty years was
to fix the so called "physical gauge" and expand perturbatively the partition
function around the classical solution (the surface of minimal area) in terms of
the inverse product of the string tension $\sigma$ with the area of the surface.

The increasing  precision of  recent numerical simulations allowed in the last
few years to test this expansion beyond the leading
order~\cite{cfghpv94,Caselle:2006dv}.
Resorting to the first order formulation
\'a la Polyakov of the NG model, it has been possible to re-sum the loop
expansion obtained in the physical gauge-fixing, in particular in the case of
Polyakov correlators \cite{Billo:2005iv} and of interfaces \cite{Billo:2006zg}.
This treatment of the NG model yields a very good agreement with
Monte Carlo results, at least for sufficiently large world-sheets; from this
point of view, the comparison with MC data sets establishes a lower bound on
the world-sheet area, below which one cannot neglect
the effects of the conformal anomaly.

Our paper lays in this context. We focus on the
universal properties of periodic interfaces arising in general 2d models. Following
the philosophy of the CWM we
make the simple assumption that the weight of the interface is
proportional to its length (which indeed corresponds to a 2d version of the  CWM)
and  treat
at the quantum level this action by means of its first-order formulation. In
this way we show in section \ref{sec:model} that the dominant term in the
partition function acquires an universal form, proportional to to $LmK_1(mR)$,
where $m$ is the inverse of the correlation length, $R$ is the lattice size in
the direction of the interface, $L$ the lattice size orthogonal to the interface
and $K_1$ is the modified Bessel function of the first order.

In section \ref{sec:ising} we consider an explicit statistical model, the 2d
Ising model. Upon a suitable choice of boundary conditions, this system allows
the formation of interfaces. The 2d Ising model is
under full analytic control, and we can derive directly the form of the
dominant term in the interface partition function. We find agreement with the
universal expression predicted by our 2d CWM.

In section \ref{sec:dimrid}, we consider the exact expression of the free energy
for interfaces in 3d obtained in \cite{Billo:2005iv}, and we perform a
dimensional reduction in one of the directions along the interface. We retrieve
in this way the 2d expression proportional to $K_1(mR)$ which we propose in
this paper. This dimensional reduction is the exact analogue for the
interface boundary conditions of the dimensional reduction from Polyakov loop
correlators to spin-spin correlators studied in~\cite{cdgjm06,cgm06}.
A non trivial consistency test of this correspondence is that the relation which
links the 2d mass $m$ with the 3d string tension $\sigma$ is the same in the
case of interfaces and of Polyakov loops.

Recently, a new set of high precision Monte Carlo data for the free energy
of interfaces in the 3d Ising model became available~\cite{chp07}. These
data include asymmetric geometries in which one of the sides of the interface
becomes much smaller than the other; in this situation one expects that the
dimensionally reduced expression can describe the data accurately. In the last
section, we test this expectation comparing these data with the predictions of
the full 3d Nambu-Goto treatment and of the 2d simple model described here.

\section{A simple model for 2d interfaces}
\label{sec:model}
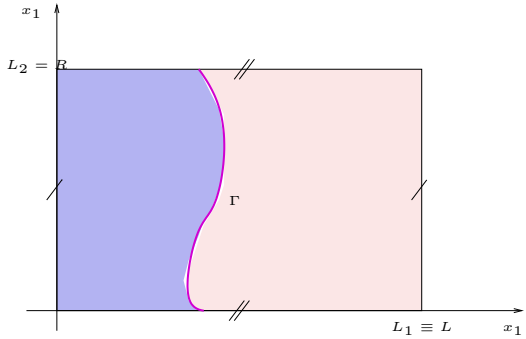
\begin{figure}
\begin{center}
\begin{picture}(0,0)%
\includegraphics{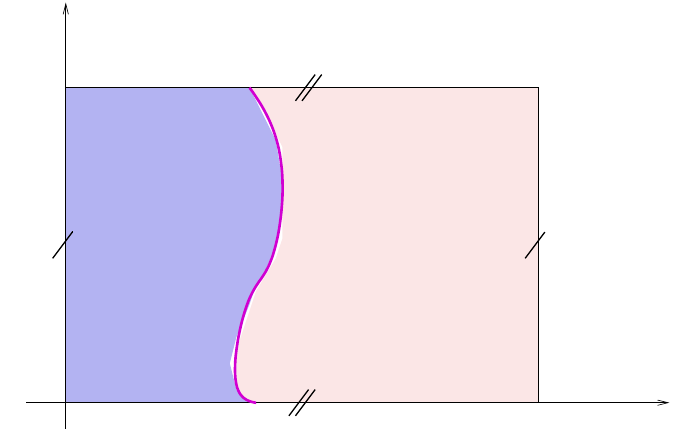}%
\end{picture}%
\setlength{\unitlength}{1658sp}%
\begingroup\makeatletter\ifx\SetFigFont\undefined%
\gdef\SetFigFont#1#2#3#4#5{%
  \reset@font\fontsize{#1}{#2pt}%
  \fontfamily{#3}\fontseries{#4}\fontshape{#5}%
  \selectfont}%
\fi\endgroup%
\begin{picture}(7811,4945)(451,-4919)
\put(6151,-4861){\makebox(0,0)[lb]{\smash{{\SetFigFont{5}{6.0}{\familydefault}{
\mddefault}{\updefault}$L_1\equiv L$}}}}
\put(7801,-4861){\makebox(0,0)[lb]{\smash{{\SetFigFont{5}{6.0}{\familydefault}{
\mddefault}{\updefault}$x_1$}}}}
\put(676,-136){\makebox(0,0)[lb]{\smash{{\SetFigFont{5}{6.0}{\familydefault}{
\mddefault}{\updefault}$x_1$}}}}
\put(451,-961){\makebox(0,0)[lb]{\smash{{\SetFigFont{5}{6.0}{\familydefault}{
\mddefault}{\updefault}$L_2=R$}}}}
\put(3751,-2986){\makebox(0,0)[lb]{\smash{{\SetFigFont{5}{6.0}{\familydefault}{
\mddefault}{\updefault}{$\Gamma$}%
}}}}
\end{picture}%
\end{center}
 \caption{An interface (the heavy magenta line $\Gamma$) separating two phases
(blue and red) on a two-dimensional torus.}
\label{fig:int}
\end{figure}

Consider a physical system, defined on a two-dimensional space with periodic
boundary conditions in both directions, in which two different phases, separated
by a one-dimensional interface, may coexist (see Fig. \ref{fig:int}).

We will describe this situation in a very simple way, analogous to the so-called
capillary wave model \cite{BLS,rw82} for two-dimensional interfaces in 3d
systems. We make the assumption that the effective action is just given by
the length of the interface $\Gamma$:
\begin{equation}
 \label{sel}
S[x] = m \int_\Gamma ds = m \int_0^1 d\tau \sqrt{(\dot x})^2~,
\end{equation}
where $\tau$ parametrizes the curve $\Gamma$, the dots denotes $\tau$
derivatives,  $(\dot x)^2 \equiv \dot x_i \dot x^i$ and $m$ is a constant%
\footnote{We can view $m$ as the ``mass'' of
a relativistic particle in an Euclidean target space (the torus)
whose world-line is the interface.}
with the dimension of an inverse length.
This is just the analogue of the conjecture which, in three dimensions, leads
to the CWM \cite{BLS,rw82}.

The corresponding partition function is given by the functional integral
\begin{equation}
 \label{pfx}
Z = \int Dx\, \ee^{-S[x]}~,
\end{equation}
and it should correspond to the interface free energy, in analogy to what
happens in the stringy description
of 3d interfaces \cite{Billo:2006zg}.

The action \eq{sel}, as well known, admits a first-order reformulation:
\begin{equation}
 \label{sel1}
S[x,e] = \frac{m}{2}\int_0^1 d\tau \left(\frac{(\dot x)ì2}{e} + e\right)~,
 \end{equation}
where $e(\tau)$ is the einbein%
\footnote{Namely one has $ds^2 = e^2(\tau)d\tau^2$},
so that the partition function can be written as
\begin{equation}
 \label{pfxe}
Z = \int De\, Dx\, \ee^{-S[x,e]}~.
\end{equation}

The action \eq{sel1} is invariant under reparametrizations of $\Gamma$:
\begin{equation}
 \label{rep}
\begin{aligned}
\delta\tau & = \epsilon~,\\
\delta e & = \dot\epsilon e + \epsilon \dot e~.
\end{aligned}
\end{equation}
We can gauge-fix this invariance%
\footnote{Notice that in the first-order formulation
we have to take into account the constraints which follow from the variation of
$S$ w.r.t. the einbein
$e$, which (in the chosen gauge) are simply $(\dot x)^2 = \lambda$.}
 by choosing $e(\tau) = \lambda = \mathrm{const}$, so that
the action becomes
\begin{equation}
 \label{acxl}
S[x,e] \to \frac{m\lambda}{2} + \frac{m}{2\lambda}\int_0^1 d\tau (\dot x)^2~.
\end{equation}
The constant $\lambda$ represents the length of the path
and is a Teichm\"uller parameter: it cannot be changed using the
reparametrizations
in \eq{rep}. As such, it must be integrated over in the partition function,
which now
reads
\begin{equation}
 \label{pfxl}
Z = \int_0^\infty d\lambda\, \ee^{-\frac{m\lambda}{2}}
\,Z_x\,
Z_{\mathrm{gh}}~.
\end{equation}
Here
\begin{equation}
 \label{Zx}
Z_x = \int Dx\, \exp\left\{-\frac{m}{2\lambda} \int_0^1 d\tau (\dot x)^2\right\}
\end{equation}
and we took into account the Faddeev-Popov ghosts $b,c$ for the chosen
gauge-fixing of the invariances in \eq{rep}.
For them we have, by standard treatment,
\begin{equation}
 \label{zgh}
Z_{\mathrm{gh}} = \int Db\, Dc \, \exp\left\{-\lambda\int_0^1 d\tau\,
b\partial_\tau c\right\}~.
\end{equation}
With the chosen periodic boundary conditions on $\tau$, it turns out
that
\begin{equation}
 \label{Zghl}
Z_{\mathrm{gh}}= \frac{1}{\lambda}~.
\end{equation}

The partition function $Z_x$ is simply that of a non-relativistic particle of
mass
$\mu = m/\lambda$, whose Hamiltonian is
\begin{equation}
 \label{Hx}
H = \frac{p^2}{2\mu}~.
\end{equation}
Since the particle moves on a two-dimensional torus of sides $L_1,L_2$, the
components of
the momentum are quantized:
\begin{equation}
 \label{qm}
p^i = \frac{2\pi n^i}{L_i}~,
\end{equation}
for $i=1,2$, with $n^i\in\mathbb{Z}$. The partition function of this system is
\begin{equation}
 \label{pfn}
\sum_{\{n_i\}} \exp\left\{-\frac{4\pi^2}{2\mu}\left(\frac{n_1^2}{L_1^2} +
\frac{n_2^2}{L_2^2} \right)\right\} =
\frac{L_1 L_2\,\mu}{2\pi}
\sum_{\{m_i\}} \exp\left\{-\frac{\mu^2}{2}\left(L_1^2 n_1^2 +
L_2^2 n_2^2 \right)\right\}~.
\end{equation}
In the second step above we performed a Poisson re-summation, which reorganizes
the sum in contributions
of sectors describing quantum fluctuations around classical solutions of
wrapping numbers $m_i$.

To describe an interface such as the one depicted in Fig. \ref{fig:int}, we set
\begin{equation}
 \label{setm}
m_1 = 0~,~~~~ m_2 = 1~,
\end{equation}
getting therefore (if we remember that $\mu = m/\lambda$)
\begin{equation}
 \label{Zxms}
Z_x = L_1 L_2 \frac{m}{2\pi\lambda} \exp\left\{-\frac{m}{2\lambda}
L_2^2\right\}~.
\end{equation}

Inserting \eq{Zghl} and \eq{Zxms} into the expression \eq{pfxl} of the total
partition function
we obtain finally%
\footnote{We use the integral
\begin{equation}
 \label{K1int}
\int_0^\infty \frac{d\lambda}{\lambda^\alpha}\ee^{-A^2\lambda -
\frac{B^2}{\lambda}} = 2 \left(\frac{A}{B}\right)^{\alpha-1} K_{\alpha-1}(2AB)~.
\end{equation}
}
\begin{equation}
 \label{Zxfe}
Z = L_1 L_2 \frac{m}{2\pi}\int_0^\infty
\frac{d\lambda}{\lambda^2}\exp\left\{-\frac{m}{2}\lambda -
\frac{m L_2^2}{2}\frac{1}{\lambda}\right\} \\
= \frac{L_1 m}{\pi} K_1(m L_2)~,
\end{equation}
where $K_1$ is the modified Bessel function of first order.

In the next section, we will consider a concrete physical system where the
interface free energy can be computed directly, namely the 2d Ising
model, and we will find that our proposed general expression \eq{Zxfe} does
indeed capture its dominant behaviour.

\section{Interfaces in the 2d Ising model}
\label{sec:ising}
Interfaces can be realized in the context of spin models,
since under peculiar boundary conditions their broken symmetry phase
allows separated but coexisting vacua of different magnetization.
Being interested in interfaces in 2d systems, it is natural to consider
the 2d Ising model, where an explicit analytical treatment is viable.

To evaluate the interface free energy of the 2d Ising model in a simple way, we
can take advantage of the fact that the continuum limit of the Ising model
is described by the Quantum Field Theory  of a free fermionic field of mass $m$.
This framework allows the calculation of the partition function with any
type of boundary conditions. Being interested in estimating the
interface free energy, we will impose periodic boundary conditions
in one direction and antiperiodic ones in the other direction. Explicit
expressions of various partition functions can be found in~\cite{km90} and
(using the $\zeta$ function regularization) in~\cite{is87}\footnote{Notice
however that in~\cite{is87} there is a sign mistake which was
later corrected in~\cite{km90}.}. It's worth pointing out that the solution
proposed in~\cite{km90} is nothing else than the continuum limit of Kaufmann
solution~\cite{k49}. Such continuum limit was performed by Ferdinand and Fischer
in~\cite{ff69} at the critical point and the result of~\cite{km90} is
essentially the extension of~\cite{ff69} outside the critical point.

Using the notations of \cite{km90}, eq.s (91)-(105), the partition function of a
fermion on a rectangular torus of sizes $L_1$ and $L_2$, which for
notational simplicity we shall henceforth denote as $L$ and $R$, is given
by the so-called ``massive fermion determinant":
\beq
\label{fermdet}
D_{\alpha,\beta}(m|L,R)=
\ee^{-\delta_\beta \pi L c_\beta(r)/6R} \prod_{n\in \mathbb{Z}+\beta}
(1-\delta_\alpha e^{-L\epsilon_n(r)/R})~.
\eeq
Here  $\alpha,\beta$ can take the values $0, 1/2$ and label the
boundary conditions in the $L$ and $R$ directions respectively:
the value $0$ corresponds to periodic b.c.s, $1/2$ to antiperiodic ones.
Moreover, $\delta_\alpha=e^{2\pi i \alpha}$ and
\beq
\frac{\epsilon_n(r)}{R}=\sqrt{m^2+(\frac{2\pi n}{R})^2}~.
\eeq
The adimensional variable $r\equiv mR$ sets the scale of the theory.

Depending on the boundary conditions, the coefficient $c_{\beta}(r)$ can
assume the following values:
\beq
c_{\frac12}(r)=\frac{6r}{\pi^2}\sum_{k=1}^\infty \frac{(-1)^{k-1}}{k}K_1(kr)
\eeq
or
\beq
c_{0}(r)=\frac{6r}{\pi^2}\sum_{k=1}^\infty \frac{1}{k}K_1(kr)~.
\eeq

The partition function of the Ising model with periodic boundary conditions in
both directions can be written as follows:
\beq
\label{Zp}
Z_{\mathrm{Ising}}(m|L,R)=
\frac12 \left( D_{\frac12,\frac12}+D_{0,\frac12}+D_{\frac12,0}-D_{0,0
} \right)~.
\eeq
If the $L$ direction is antiperiodic, the partition function becomes
\beq
\label{Zap}
Z_{\mathrm{Ising,ap}}(m|L,R)=
\frac12 \left( D_{\frac12,\frac12}+D_{0,\frac12}
-D_ { \frac12 , 0 } -D_ { 0 ,0}\right)~.
\eeq
This is the situation which generates an odd number of interfaces along the $R$
direction. The interface free energy will be given by the
following expression:
\beq
\label{Fising}
\ee^{-F_{\mathrm{interface}}} \equiv Z_{\mathrm{interface}}
= \frac{Z_{\mathrm{Ising,ap}}}{Z_{\mathrm{Ising}}}~.
\eeq

Notice, as a side remark, that taking the limit%
\footnote{In this regime, one must keep in mind that the parameter $\tau$ of
\cite{ff69} corresponds to $r/2$.}
$r\to 0$, one moves to the critical point and all the expressions written above
flow toward the conformal invariant ones and agree with the standard CFT results
 and with those reported in the  pioneering work of
 Ferdinand and Fisher \cite{ff69}.

We are now interested in the large $mL$ and $mR$ limit of eq.(\ref{Fising}).
The large $mL$ limit allows us to neglect the infinite product in
eq.(\ref{fermdet}) while in the large $mR$ we approximate
\beq
c_{\frac12}(r)\sim c_0(r)\sim \frac{6r}{\pi^2}K_1(r)~.
\eeq
The  partition functions in eq.(\ref{Zp}) and (\ref{Zap}) simplify and we
end up with
\beq
\begin{aligned}
Z_{\mathrm{Ising}} & \sim 1~,
\\
Z_{\mathrm{Ising,ap}} & \sim \frac{\pi L}{6R} \left( c_{\frac12}(r) + c_{0}(r)
\right)~,
\end{aligned}
\eeq
from which we find
\beq
\label{Zisint}
Z_{\mathrm{interface}}\sim \frac{\pi L}{6R} ( c_{\frac12}(r) + c_{0}(r) )
\sim\frac{2 L r}{\pi R} K_1(r) =\frac{2 L m}{\pi } K_1(mR)~.
\eeq
By comparing this result to \eq{Zxfe}, we see that the behaviour of the
interface free energy in the 2d Ising model for large scales is perfectly
captured by our simple effective model.

\section{Dimensional reduction of the Nambu-Goto effective description of 3d
interfaces}
\label{sec:dimrid}
In this section we shall discuss the behaviour of the Nambu-Goto effective
string description of
fluctuating interfaces when dimensional reduction occurs along
one of the two lattice sizes which define the interface.

As we discussed above, interfaces can be realized in spin
models. In this realization, the physical ideas underlying the
dimensional reduction process are perhaps best discussed. Focusing on the 3d
Ising model, let us recall that it is mapped by duality into the 3d Ising gauge
model. Under this mapping, the interface free energy becomes a close
relative of the Wilson loop expectation value of the 3d gauge Ising model, the
only difference being in the boundary conditions of the two observable: fixed in
the Wilson loop case and periodic in the interface case.

In lattice gauge theories, dimensional reduction has a very important physical
meaning. The size of the lattice direction which is reduced is interpreted as
the inverse temperature of the system and the point in which dimensional
reduction occurs coincides with the deconfinement phase transition. The
behaviour of the model in the vicinity of the deconfinement transition is well
described, according to the Svetitsky-Yaffe conjecture \cite{sy82}, by the 2d Ising
spin model. More generally, for other gauge groups with continuous
deconfinement phase transitions, it is described by the 2d spin model with
global symmetry group the center of the original gauge group.

As a consequence of this observation, we expect that the interface free
energy of the 3d Ising model
 should smoothly map into the 2d one when the dimensional reduction scale
is approached from above, while it should vanish below it, since in the dual
model this is the deconfined phase. Any reliable effective description of the
free energy should therefore be compatible with these two requirements. This is
a rather non trivial test for the effective string models which are expected to
effectively describe both the Wilson and Polyakov loop expectation value and the
interface free energy.

In this section, we start from the expression for the interface
free energy in 3d obtained in \cite{Billo:2006zg} and consider its behaviour
under dimensional reduction; we
find that it does indeed smoothly reduce to the 2d expression given here in
\eq{Zxfe}. This represents a test of our effective approach to the interface
free energy both in three and in two dimensions.

In the next section, we will refine this test by comparing both the 3d and the
2d theoretical expressions to Monte Carlo data for asymmetric interfaces: the
more asymmetric are the interfaces, the more we expect the prediction of our 3d
and 2d expression to become compatible and reliable.

Let us consider a three-dimensional torus of sides $L_{1,2,3}$, and assume that
the interface lies in the plane orthogonal to $L_1$.
Our starting point is the expression for the 3d interface partition function in
the Nambu-Goto approximation
 evaluated in \cite{Billo:2006zg} (eq.s (2.23, 2.24)):
\begin{equation}
\label{I3}
\mathcal{I}^{(3)} = 2 \left(\frac{\sigma}{2\pi}\right)^ {\frac{1}{2}}\, L_1 \,
\sqrt{\sigma\mathcal{A}u}
\sum_{k,k'=0}^\infty  c_k c_{k'}
\left(\frac{\cale}{u}\right)  K_{1}\left(\sigma\mathcal{A}\cale\right)~,
\end{equation}
where
\begin{equation}
\mathcal{A}=L_3 L_2~,~u=\frac{L_2}{L_3}
\end{equation}
and
\begin{equation}
\cale  =
\sqrt{1 + \frac{4\pi\, u}{\sigma \mathcal{A}}(k+k'-\frac{1}{12}) +
\frac{4\pi^{2} \, u^2\, (k-k')^{2}}{(\sigma\mathcal{A})^2}}
\end{equation}
is a function of the occupation numbers $k$ and $k'$.
The coefficients $c_k$ and $c_{k'}$  are the number of partitions of $k$ and
$k'$.

The dimensional reduction of \eq{I3} can be most easily discussed by writing the
interface free energy as a function of $L_3$. Following the notations of
\cite{Billo:2006zg}, let us introduce the ``energy levels'' $E_{k,k'} \equiv
\sigma L_3 \cale$. With this choice the argument of the Bessel functions in
eq.(\ref{I3}) becomes $L_2 E_{k,k'}$ and we have:
\begin{equation}
\label{spectrum}
E_{k,k'} = \sqrt{\sigma^2 L_3^2 + 4\pi\sigma (k+k'-\frac{1}{12}) +
\frac{4\pi^{2}\, (k-k')^{2}}{L_{3}^{2}}}~.
\end{equation}
From this expression it is evident that, as the ratio $L_2/L_3$ increases, the
separation between consecutive levels in the spectrum becomes larger and larger
and, in particular, the gap between the lowest state $k=k'=0$ and the second one
increases.

In the dimensional reduction limit $L_3\ll L_2$ we can thus truncate the sum
in eq.(\ref{I3}) to the first term. In this limit, the argument of the
Bessel function in eq.(\ref{I3})
becomes large,  allowing us to use the asymptotic expansion
\begin{equation}
\label{bessel}
K_j(z) \sim \sqrt{\frac{\pi}{2z}}e^{-z}\left\{1+O\left(\frac{1}{z}\right)\right\},
\end{equation}
from which we see that higher states are exponentially suppressed with respect
to the $k=k'=0$ one. Setting $k=k'=0$, we get
\begin{equation}
\label{I3_00}
\mathcal{I}^{(3)}_{(0,0)} = 2 \left(\frac{\sigma}{2\pi}\right)^ {\frac{1}{2}}\,
L_1 \, \sqrt{\sigma\mathcal{A}u}
\left(\frac{\cale}{u}\right)  K_{1}\left(\sigma\mathcal{A}\cale\right)~,
\end{equation}
where now
\begin{equation}
\label{cale00}
\cale  =
\sqrt{1 - \frac{\pi}{3\sigma \mathcal{A}}u}=\sqrt{1 - \frac{\pi}{3\sigma L_3^{2}}}~.
\end{equation}
Equation (\ref{I3_00}) can therefore be rewritten as:
\begin{equation}
\mathcal{I}^{(3)}_ {(0,0)} =  \left(\frac{2}{\pi}\right)^ {\frac{1}{2}}\, L_1 \, \sigma L_3
 \sqrt{1 - \frac{\pi}{3\sigma L_3^{2}}} K_{1}\left(\sigma L_3 L_2\sqrt{1 - \frac{\pi}{3\sigma
 L_3^{2}}}\right)~.
\end{equation}
If we now  define
\begin{equation}
\label{meff}
m_{\mathrm{eff}} = \sigma L_3 \sqrt{1 - \frac{\pi}{3\sigma L_3^{2}}}
\end{equation}
we find
\begin{equation}
\mathcal{I}^{(3)}_ {(0,0)} =  \left(\frac{2}{\pi}\right)^ {\frac{1}{2}}\, L_1
\, m_{\mathrm{eff}} K_{1}\left(m_{\mathrm{eff}} L_2\right)~,
\end{equation}
which is in perfect agreement with the partition function given by (\ref{Zxfe})
a part from the normalization constant.

It is clear from \eq{cale00} that the above discussion is consistent
only for
\begin{equation}
L_3~\geq~\sqrt{\frac{\pi}{3\sigma}}~.
\end{equation}
In the string language, this bound represents the tachyonic singularity of the
bosonic string: for $L_3~\leq~\sqrt{\frac{\pi}{3\sigma}}$ the lowest state has a
negative energy. In the 3d Ising model framework, the critical value
\begin{equation}
\label{lc}
 L_{3,\mathrm{c}}= \sqrt{\frac{\pi}{3\sigma}}=1.023...
\times\frac{1}{\sqrt{\sigma}}
\end{equation}
at which the mass of the lowest state vanishes can be considered as the
effective string prediction for the dimensional reduction scale which we were
looking for the dimensional reduction scale which we were looking
for. We thus see that this scale is naturally embedded in the NG formulation of
the interface
free energy. The same argument in the dual case of the Polyakov loop correlator
leads to the identification of this scale with deconfinement temperature
\cite{olesen}.
The value obtained in this way for the dimensional reduction
scale (or, equivalently, for the deconfinement temperature in the dual model) is
in rather good agreement with the Monte Carlo estimates. For the 3d gauge Ising
model the disagreement is only of about $20\%$, as the Monte Carlo estimate
turns out to be $\sqrt{\sigma}L_c=0.8186(16)$~\cite{ch96}  instead of 1.023...)
and it further decreases if one looks at $SU(N)$ pure Yang Mills theories.

It is important to stress that the mapping between 3d and 2d observables which
emerges from this comparison is exactly the same which is found considering the
finite temperature behaviour of the Polyakov loop correlators in the vicinity
of the deconfinement transition in the
3d gauge Ising model. In particular, the
relation between the mass scale of the 2d model and the string tension (in the
present case the interface tension) of the 3d one reported in eq. (\ref{meff})
is the same which one finds in the Polyakov loop case. This is a
rather non trivial consistency check of the whole dimensional reduction picture.

As a last remark let us stress that the smooth flow of the 3d Nambu-Goto result
towards the 2d ones under dimensional reduction has various important
implications on our understanding of the Nambu-Goto effective action. Recall
that the treatment leading to eq. (\ref{I3}) neglected the conformal anomaly,
i.e., equivalently, did not take into account the Liouville field which does not
decouple in the Polyakov formulation of the string when $d\not=26$. Most
probably
the smooth flow toward the 2d result occurs exactly because we neglected the
Liouville field contribution. It is thus interesting to compare simultaneously
the Nambu-Goto expression (without Liouville field), the 2d result presented
here and the output of the Montecarlo simulations for the free energy of
interfaces on lattices with a value of $L_3$ small enough to make the 2d
approximation a reliable one. We shall devote the next section to this
comparison.

\section{Comparison with the numerical data}
\label{sec:num}
Following the above discussion we compare in this section
the expressions for the 2d interface free energy given
in \eq{Zxfe} and for the 3d free energy \cite{Billo:2006zg}, reported in
\eq{I3}, with a set of high-precision Monte Carlo data
\cite{chp07} obtained in a 3d Ising model in which interfaces are created by
imposing  antiperiodic boundary conditions in one of the three lattice
directions.

In particular we isolated two sets of data. The first one is reported in Tab.
\ref{tab16} and displays the values of the free energies obtained in the 3d
Ising model at $\beta = 0.223101$. The interface string tension for this value
of $\beta$ is $\sigma = 0.0026083$ and the dimensional reduction scale is
exactly $16$ lattice spacings~\cite{ch96}. The data  correspond to values of the
$L_3$ size ranging from a minimum value of $3/2$ the dimensional reduction scale
up to $3$ times.

In the table we report, from left to right, the lattice sizes $L_{1,2,3}$ , the
numerical estimate $F_{\mathrm{num}}$ for the free energy (with its statistical
error), the theoretical estimate $F$ according to \eq{I3}, the estimate
$F_{(0,0)}$ obtained in the full dimensional reduction limit by truncating
\eq{I3} to $k=k'=0$, which corresponds to our 2d formula \eq{Zxfe},
and the estimate $F_{1\mathrm{st}}$ in which also the first excited states,
$k+k'=1$, are kept into account. Finally, as a comparison, in the last two
columns we report the one loop and two loop perturbative approximations to the
whole NG action discussed, for instance, in \cite{df83,cfghpv94}.

In Tab. \ref{tab8} we report the same set of data evaluated at $\beta =
0.226102$, corresponding to $\sigma = 0.0105254$ and to a dimensional reduction
scale of $8$ lattice spacings~\cite{ch96}.

\begin{table}
\begin{center}
\begin{tabular}{cccccccccc}
\hline
$L_3$ & $L_2$ & $L_1$ & $F_{\mathrm{num}}$ & $F$ & $F_{(0,0)}$ &
$F_{1\mathrm{st}}$ & $F_{\mathrm{1-loop}}$ & $F_{\mathrm{2-loop}}$ \\
\hline
24 & 64 & 96 & 6.8855(20) & 6.7495 & 6.7495 & 6.7495 & 6.9974 & 6.8347\\
28 & 64 & 96 & 7.6929(21) & 7.6537 & 7.6537 & 7.6537 & 7.7875 & 7.6821\\
32 & 64 & 96 & 8.4626(20) & 8.4518 & 8.4518 & 8.4518 & 8.5380 & 8.4632\\
36 & 64 & 96 & 9.1996(21) & 9.2012 & 9.2012 & 9.2012 & 9.2632 & 9.2062\\
40 & 64 & 96 & 9.9227(23) & 9.9231 & 9.9232 & 9.9231 & 9.9713 & 9.9253\\
44 & 64 & 96 & 10.6203(23) & 10.6278 & 10.6280 & 10.6278 & 10.6674 & 10.6288\\
48 & 64 & 96 & 11.3138(25) & 11.3209 & 11.3214 & 11.3209 & 11.3548 & 11.3213\\
\hline
\end{tabular}
\end{center}
\caption{Numerical results of the free energy in the 3d
Ising model at $\beta = 0.223101$ are compared to various theoretical
estimates. See the main text for detailed explanations.}
\label{tab16}
\end{table}

\begin{table}
\begin{center}
\begin{tabular}{cccccccccc}
\hline
$L_3$ & $L_2$ & $L_1$ & $F_{\mathrm{num}}$ & $F$ & $F_{(0,0)}$ &
$F_{1\mathrm{st}}$ & $F_{\mathrm{1-loop}}$ & $F_{\mathrm{2-loop}}$ \\
\hline
24 & 64 & 96 & 18.4131(26) & 18.4121 & 18.4121 & 18.4121 & 18.4555 & 18.4152\\
28 & 64 & 96 & 21.2414(27) & 21.2450 & 21.2450 &  21.245 & 21.2724 & 21.2463\\
32 & 64 & 96 & 24.0310(27) & 24.0306 & 24.0306 & 24.0306 & 24.0497 & 24.0312\\
36 & 64 & 96 & 26.7859(28) & 26.7873 & 26.7873 &  26.7873& 26.8017 & 26.7875\\
40 & 64 & 96 & 29.5271(30) & 29.5250 & 29.5251 & 29.5250 & 29.5365 & 29.5251\\
44 & 64 & 96 & 32.2449(31) & 32.2498 & 32.2500 & 32.2498 & 32.2594 & 32.2498\\
48 & 64 & 96 & 34.9623(33) & 34.9653 & 34.9657 & 34.9653 & 34.9736 & 34.9653\\
\hline
\end{tabular}
\end{center}
\caption{Numerical results of the free energy in the 3d
Ising model at $\beta = 0.226102$ are compared to various theoretical
estimates. See the main text for detailed explanations.}
\label{tab8}
\end{table}

Looking at these tables one can see that the 2d expression $F_{(0,0)}$ always
gives a very good approximation of the whole NG result $F$: the difference
between the two is always smaller than the statistical error on the numerical
data. One can also notice that the 2d approximation is more reliable when the
dimensional reduction scale $\sqrt{\pi/(3\sigma)}$ is smaller, i.e. in Tab.
\ref{tab8}, and that within each data set it becomes slightly worse as $L_3$
increases. Looking at the seventh column we see that the deviation from the full
NG result is completely due to the first excited state.

In fact it is easy to evaluate the gap $\Delta$ between the lowest state and the
first excitation which appears at the exponent of the asymptotic expansion of
the Bessel function in \eq{I3}. In the large $\sigma L_3^2$ limit, the first
excited states correspond to $k=1$, $k'=0$ or $k=0$, $k'=1$ and
$\Delta$ takes the value $2\pi L_2/L_3$. When  $\pi/3 \leqslant \sigma
L_3^2 \leqslant \pi$ the first excited state is instead the one with $k=1$,
$k'=1$ and the variation of the gap is
\beq
\sqrt{\frac{8}{3}}\pi\frac{L_2}{L_3} \geqslant \Delta
\geqslant \pi\frac{L_2}{L_3}~.
\eeq
We see that the dominating effect is due to the asymmetry ratio $L_2/L_3$
(which in our data ranges from $8/3$ to $4/3$) and that a subdominant role is
played by the $\sigma L_3^2$ combination, in complete agreement with the results
reported in Table \ref{tab16} and \ref{tab8}.

Given this agreement between 2d and whole NG results it is very interesting to
compare them both to the Montecarlo data. The agreement is rather good for large
values of $F$, see in particular the results reported in Tab. \ref{tab8}. As $F$
decreases, however, the NG predictions compare less favorably to the data, see
Table \ref{tab16}. As a matter of fact, in this regime the Monte Carlo data
seem to better agree with the two loop perturbative expansion than with the
whole NG result (and its 2d approximation). This fact has already been discussed
in \cite{chp07} and is most probably due to the fact that the two loop result is
actually consistent also at the quantum level, in the sense that at this order
the contribution due to the Liouville field vanishes
\cite{ps91,Luscher:2004ib,d04,hdm06}.

As expected, these problems  disappear
if one works exactly in two dimensions, and in fact the direct calculation of
the interface partition function in the 2d Ising model reported in sec.
\ref{sec:ising} shows that the expression \eq{Zxfe} obtained in section
\ref{sec:model} should be valid to all orders.

\section{Conclusions}
In this paper, we addressed the description of some universal properties of
interfaces in two-dimensional physical systems. Using a simple model in which
one assumes an action proportional to the length of the interface, in analogy
with the 3d capillary wave model, we proposed a general expression for the
interface free energy. This expression agrees with an exact calculation in the
case of the 2d Ising model. We also shew that our 2d result represents the
dimensional reduction of the interface 3d free energy obtained using the
capillary wave model, which is tantamount to the Nambu-Goto effective string
description, when the effects of the conformal anomaly are neglected. Finally,
we compared the 3d and 2d theoretical predictions to Monte Carlo numerical
estimates for the interface free energy in the case of the 3d Ising model.
In the range of values that we studied the difference between the 3d and 2d
estimates is always smaller than the statistical error of the numerical data.
Moreover we were able to show that this difference is completely due to the
first excited state of the Nambu-Goto action.

\acknowledgments{We warmly thank F. Bastianelli, O. Corradini, F. Gliozzi,
M.Hasenbusch and M. Panero for useful discussions.}

\end{document}